\documentclass[fleqn,10pt]{SelfArx} 

\usepackage[english]{babel} 

\usepackage{lipsum} 
\usepackage{colortbl}
\usepackage{multicol}
\usepackage{algpseudocode,algorithm,algorithmicx}
\usepackage{amsmath}
\usepackage{caption} 
\usepackage{subcaption} 
\usepackage{graphicx}

\newcommand{\savefootnote}[2]{\footnote{\label{#1}#2}}
\newcommand{\repeatfootnote}[1]{\textsuperscript{\ref{#1}}}

\definecolor{color1}{RGB}{0,0,90} 
\definecolor{color2}{RGB}{0,20,20} 

\usepackage{hyperref} 

\hypersetup{
	hidelinks,
	colorlinks,
	breaklinks=true,
	urlcolor=color2,
	citecolor=color1,
	linkcolor=color1,
	bookmarksopen=false,
	pdftitle={Title},
	pdfauthor={Author},
}

\JournalInfo{Draft Version} 
\Archive{May, 2023} 

\PaperTitle{Open Framework for Analyzing Public Parliaments Data} 

\Authors{Shai Berkovitz\textsuperscript{*}, Amit Mazuz\textsuperscript{*}, and Michael Fire} 
\affiliation{\textit{Department of Software and Information Systems Engineering, Ben-Gurion University, Israel\\
\{shaiber,amitmazu,mickyfi\}@post.bgu.ac.il}\\
*Authors contributed equally} 

\newcommand{\keywordname}{Keywords} 

\Abstract{Open information about government organizations should interest all citizens who care about their governments' functionality. Large-scale open governmental data open new opportunities for citizens and researchers to monitor their government's activities and improve its transparency. Over the years, various projects and systems have processed and analyzed governmental data based on open government information. Here, we present the Collecting and Analyzing Parliament Data (CAPD) framework. This novel generic open framework enables collecting and analyzing large-scale public governmental data from multiple sources. We used the framework to collect over 64,000 parliaments protocols from over 90 committees from three countries. Then, we parsed the collected data and calculated structured features from it. Next, using the calculated features, we utilized anomaly detection and time series analysis to uncover various insights into the committees' activities. We demonstrate that the CAPD framework can be utilized to identify anomalous meetings and detect dates of events that affect the parliaments' functionality and help to monitor their activities. }

\begin{document}

\maketitle 


\thispagestyle{empty} 


\section{Introduction} 
Open information about government organizations should interest all citizens who care about their governments' functionality. According to Hulstijn et al ~\cite{hulstijn2017open}, four key issues are essential for open information: government transparency, improving public service, innovation, economic value, and efficiency. Moreover, open information is essential in various fields, such as health, emergency, transportation, etc.

In addition, public data is a crucial source of social innovation and economic growth. Open data provides new opportunities for governments to collaborate, as many do at innovation events and competitions. Businesses and entrepreneurs use open data to understand potential markets better and build data-driven products \cite{kitsios2018open}.

Over the years, various projects and systems have processed and analyzed governmental data using open government information. One type of project focuses on analyzing public data-sets data, such as the \textit{Openkamer}\footnote{\url{https://github.com/openkamer/openkamer}} project, operating in Dutch and provides insights into the Dutch parliament. Other projects focus on collecting and processing government data by public APIs or web scrapping and making it more accessible to the public. For example, the \textit{Open Knesset}\footnote{\url{https://oknesset.org}} project mines the Israel parliament (Knesset) activities from the official Knesset website to track voting, legislation, and committee activities.

\begin{figure*}[!ht]
    \centering
    \scalebox{0.30}{\includegraphics{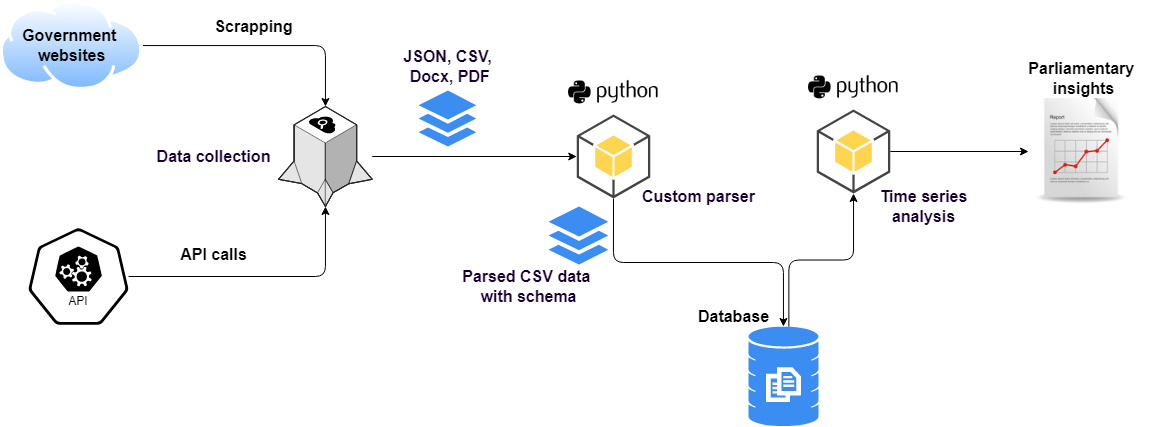}}
    \caption{Framework Overview.}
    \label{fig:framework_flow}
\end{figure*}

This study presents the Collecting and Analyzing Parliament Data framework (CAPD framework). This generic novel framework offers data analysis of government and parliament information presented as time series. This lets us get insights and anomalies about the organization's work habits and functionality during certain events or periods. Our framework is built from three primary components (see Figure~\ref{fig:framework_flow}): The first component collects parliamentary data from different countries using public APIs and custom web scrappers. The second component parses the collected data and extracts features from it. The last part performs a time series analysis of the extracted features and generates insights into the parliament's functionality in a specific period. We evaluated our framework's ability to identify events and changes in parliament committees' activities by collecting large-scale parliamentary data from the USA, Canada, and Israel. Moreover, using the framework, we analyzed the curated data using machine learning algorithms (see Section~\ref{sec:methods_and_experiments}). The results show that the open CAPD framework can collect and analyze a large amount of data and detect dates of events that affect parliaments' functionality, like COVID-19 days and special discussions in specific committees (see Section~\ref{sec:results}).

This paper presents three key contributions:
\begin{itemize}
    \item We introduce a novel open generic framework that can be used to collect and analyze large-scale parliamentary data from different parliaments worldwide. The framework can be utilized to collect and analyze additional parliamentary data and advance data-driven research of governments' activities. Moreover, the framework can be used to monitor parliament activities.
    
    \item We curated a large-scale open data set with over 64,000 protocols collected from different countries.
    
    \item We demonstrate that using Change Point Detection algorithms (see Section~\ref{sec:timeseries}) can be used to analyze parliamentary data and detect interesting committee meetings with relatively high precision (see Section~\ref{sec:results}).
    
\end{itemize}

The remainder of this article is organized as follows. Section \ref{sec:related_work} overviews open government information publicity and its contribution to government activity. Additionally, we examine the related systems that analyze government information and give some knowledge about the time series analysis used in our research.

Next, in Section~\ref{sec:methods_and_experiments}, we describe our methodology for building our framework for detecting change points and anomalies in parliamentary data and our way of evaluating the various methods. Afterward, in Section~\ref{sec:results}, we will represent the different results of our collected data and analyzing methods and compare the different analysis algorithms. Subsequently, in Section~\ref{sec:diss}, we discuss the results. Lastly, Section~\ref{sec:conclusions} presents the study conclusions and future research directions.

\section{Related Work}
\label{sec:related_work}

This Section will provide an overview of related works to our multidisciplinary study. First, in Section ~\ref{sec:open-goverment-info}, we will review and detail the disclosure of information in different governments and parliaments worldwide. Afterward, in Section~\ref{sec:contb-goverment-activity}, we describe the contributions open information in governments can make. Then, in Section~\ref{sec:systems_gov}, we review the various parliament systems and what they can offer the public. Subsequently, in Section~\ref{sec:related_datamining}, we provide a short overview of studies in which data mining algorithms analyze open government data (OGD). Lastly, in Section~\ref{sec:timeseries}, we will present several methods for time series analysis used in this study.

\subsection{Open Government Information} \label{sec:open-goverment-info}

Open government policy, as it has developed in the Western world in recent years, seeks to harness new technologies for the benefit of improving the means of communication between government and its citizens and for the proper utilization of the social and economic benefits inherent in the information held by government authorities~\cite{fetisova2020role}. Public sector information technologies are used at the federal and regional levels to improve the efficiency of government bodies' administrative activities /cite[fetisova2020role].

One of the main benefits of promoting transparency, access to information, and public participation in the digital age is the ability to monitor the government's actions~\cite{guggisberg2021transparency}. Another advantage is that access to information in various fields makes it possible to derive public and economic benefits and encourage development and entrepreneurship. Of course, public participation is an excellent tool for implementing this concept and principle of good governance. It contributes to responsibility and enables public participation, the collective perception of a group of people, not necessarily experts~\cite{guggisberg2021transparency}.

Additionally, according to Barbra Ubaldi~\cite{ubaldi2013open}, governments must seek feedback from the public on the usefulness, relevance, and accessibility of their data to enable continuous improvement. Furthermore, many civil servants see the real-time performance and impact of public services and public policy on citizens. They can create appropriate data and other inputs, or use available data, to improve the service experience, provided the tools and incentives are provided, for example, by providing the opportunity to participate in a professional role on online social networks to offer advice and information to the public~\cite{ubaldi2013open}. Therefore, governments need to recognize the value of the audience source to enable real-time data and information sharing and engage relevant stakeholders outside public organizations to use them to create value~\cite{ubaldi2013open}.

Many governmental datasets are accessible to the public by several interfaces: (a) \textit{requesting information} on relevant government websites; (b) consuming the information \textit{through APIs}\savefootnote{api-footnote}{Application Programming Interface (API) is a set of code libraries or pre-made functions, which programmers can make easy use of, without having to write them themselves so that they can use the information of the application they want to use for their application.} provided by the government to the public; or (c) by \textit{crawling and scraping} various government websites. Here are some examples:

\textbf{Apply online - FOIA.} There is a UK website\savefootnote{foia-uk-footnote}{\url{https://www.foia.gov/}.} detailing how to apply; The request for information can be submitted by sending a written request to any public body. The body must return an answer within 20 days. Most applications are free of charge. However, if there are expenses associated with the required body, it may request payment for this request for information. This is before performing the work. In the US and the UK, you must identify the body (agency) to which the application should be sent (each body handles the request). It is important to note that the information seeker pays for search hours. The public body commits to deadlines according to the body and information requested. If the request is too extensive, the body can reject it.

\textbf{Using API access.} In the UK, the API approach is currently in the ALPHA stages.\savefootnote{uk-apis-footnote}{\url{https://www.api.gov.uk/}.} However, it offers about 200 access points (URL endpoints) on 26 topics. Each access point contains documentation and an API. In the US, access to information through the API is quite advanced and high-quality. The API offers various data from 40 access points on 18 topics.

\textbf{Crawling and Scrapping online data.} Data scraping is the process of extracting data from a human-readable output from another program. Web crawling is a type of data scraping that aims to explore the web, primarily finding URLs and links. Web scraping is also a type of data scraping that extracts data and information from web-pages~\cite{khder2021web}. In the USA, you can crawl for links to all the committee's pages, and afterward, you can scrape valuable data like congressional bills, reports, hearings, and prints.\savefootnote{us-govinfo-footnote}{\url{https://www.govinfo.gov/browse/committee}.} In Israel, for example, you can scrape all the Knesset protocols by writing a scrapper to a single web-page.\savefootnote{il-knesset-scrapping-footnote}{\url{https://main.knesset.gov.il/Activity/committees/Pages/AllCommitteeProtocols.aspx}.}

In recent years, several federal data monitoring and conservation projects have been conducted that are scientifically hosted on government databases in the US and other countries~\cite{carrara2015creating}.

Today, governments are trying to be more ``open.''\savefootnote{ips_report}{\url{https://www.ipc.nsw.gov.au/media/3241}} One aspect of open government is opening governmental data~\cite{janssen2012benefits,martin2013open,zuiderwijk2012socio}. However, it is known that simply providing open government data does not automatically produce significant value for society~\cite{janssen2012benefits}. The literature often cites the many potential benefits of OGD~\cite{janssen2012benefits,foulonneau2014open,janssen2011influence,ubaldi2016rebooting}. However, it is still believed that these benefits will not materialize unless the data is used. Thus, a concrete understanding of the barriers that prevent OGD use to produce public value is essential. As a continuation of this, a framework is needed that will guide the use of OGD efficiently and effectively that produces as much public value as possible~\cite{foulonneau2014open}.

\subsection{Contribution to Government Activity}
\label{sec:contb-goverment-activity}

Governments collect data for future use~\cite{gonzalez2021open}. For example, during the COVID-19 pandemic, researchers and government employees used existing information~\cite{gonzalez2021open}. They have increased the demand for timely, relevant, and quality data access. Several needs drive this demand: taking informed and rapid policy action, improving engagements, conducting a scientific analysis of a dynamic threat, and understanding social and economic impacts that enable oversight and reporting on Civil Society. Data and AI are components of innovation that can help us find solutions to social challenges, from health to agriculture and security to manufacturing \cite{gonzalez2021open}. On the other hand, the multiple uses of data must be balanced while maintaining high standards of privacy, security, safety, and ethics \cite{huyer2020economic}.

For example, the World Health Organization (WHO) stated that they used open data and location analytics to eliminate malaria in their annual report.\savefootnote{malaria-report-footnote}{\url{https://apps.who.int/iris/bitstream/handle/10665/275867/9789241565653-eng.pdf?ua=1}.} Thousands of volunteers have worked on mapping hundreds of thousands of square miles of the Malaria-affected world, leading to a more efficient distribution of resources and identification of much-needed areas. The collection and use of improved data contributed to an 85\% decrease in reported Malaria cases and a 92\% decrease in Malaria-related deaths across the southern province, affecting approximately 1.8 million people.\repeatfootnote{malaria-report-footnote}

It is critical to note that researchers can use government data to draw groundbreaking conclusions. This occurred during the treatment of the COVID-19 pandemic~\cite{sohrabi2020world}. Although large amounts of government data have been made available through databases and portals, there needs to be more evidence of dedicated services or innovations created from OGD reuse. In line with broader calls to ensure a ``targeted advertising'' approach, organizations need to understand better the needs of data across the ecosystem~\cite{gonzalez2021open}.

In 2015, it was expected that using OGD would save European governments 1.7 billion euros and produce 350,000 jobs by 2020~\cite{carrara2015creating}. These goals have been achieved, as seen in a report published in the European Data Portal in 2020.\savefootnote{data-portal-report-footnote}{\url{https://data.europa.eu/sites/default/files/sustainability-data-portal-infrastructure_6_distributed-version-control.pdf}.}  One of the reasons for achieving these goals is the number of resources Europe invests in open-data, especially in OGD. In fact, except for one parliament, all parliaments in Europe have a body whose whole purpose is to promote the use and quality of OGD~\cite{huyer2020economic}. These bodies coordinate open data initiatives and create supporting materials to publish and reuse available data. They provide training and workshops to civil servants and promote open data publication and reuse.

\subsection{Systems for Analyzing Government Information}
\label{sec:systems_gov}

In recent years, more bodies, public institutions, and governments have acted accordingly and adapting to this. Today, numerous data sets are offered to the general public in many countries, including more relevant and usable information. 

For example, the CA and the US parliaments’ portals contain over 25,000\savefootnote{data-ca}{\url{https://open.canada.ca/en/open-data}} and 330,000\savefootnote{data-gov}{\url{https://data.gov}} data-sets, respectively. 

In addition, there have been different open code initiatives utilizing available governmental data sets. The \textit{Openkamer} code project\footnote{\url{https://github.com/openkamer/openkamer}} provides insight into the Dutch parliament by extracting parliamentary data from several external sources. It visualizes this data in a web application, such as legislative proposals, queries, political parties, gifts to parliament, and more. Similarly, in the UK, the Public Whip project,\footnote{\url{https://www.publicwhip.org.uk}}  an independent non-governmental project, web-scrapes House of Commons and House of Lords debate transcripts to enable the public to monitor and influence voting patterns.\footnote{\url{https://www.publicwhip.org.uk/faq.php}} Additionally, in France and Canada, similar projects have been implemented. In France, the \textit{senapy} Python project\footnote{\url{https://github.com/regardscitoyens/senapy}} scrapes data from the French Senate website.\footnote{\url{https://senat.fr}} Whereas in Canada, the \textit{coalition-analyzer}\footnote{\url{https://github.com/oliversno/coalition-analyzer}} code project submits open API requests for the Canadian House of Representatives to calculate correlations in the votes of different parties. In Israel, the \textit{Open Knesset}\footnote{\url{https://oknesset.org}} project mines all Israel parliament activity from the official Knesset website to track voting, legislation, and committee activities.

\subsection{Data Mining Parliamentary Data}
\label{sec:related_datamining}

Several studies took public parliamentary information and applied natural language processing (NLP) methods to detect political ideology, sentiment, position‑taking, perspectives, and the level of agreement and disagreement between politicians.

In 2003, Laver and Bendit~\cite{Laver2003ExtractingPP} presented a method for extracting policy positions from political text. They start by implementing their method and testing it on Britain and Ireland's political parties. Then, they ``export'' their model to non-English languages like German and extend the model from party manifesto analysis to estimate political positions from legislative speeches.

In 2012, Awadallah et al.~\cite{awadallah2012opinions} presented \textit{OpinioNetIt}, a structured, faceted knowledge-base of opinions, and its use in political views analysis.

They focused on acquiring opinions held by various stakeholders on politically controversial topics. Awadallah et al.'s system can be used for multiple types of analysis, including heatmaps showing political bias, flip-flopping politicians, dissenters, and more~\cite{awadallah2012opinions}.

In 2014, Iyyer et al.~\cite{iyyer-etal-2014-political} implemented a Recurrent Neural Network (RNN) for identifying the political position evinced by a sentence. They used multiple public data sets containing over a million sentences. Then, they filtered, annotated, and processed them to adjust their identification task, resulting in almost 100,000 labeled sentences.

In 2017, Vilares proposed ``a Bayesian modeling approach where topics (or propositions) and their associated perspectives (or viewpoints) are modeled as latent variables''~\cite{vilares2017detecting}. They evaluated their model on debates from the House of Commons of the UK Parliament scrapped using a custom web crawler. They revealed perspectives from debates without labeled data. In the same year, Gencheva et al.~\cite{gencheva2017context} constructed an entire corpus of political debates containing statements that reputable sources fact-check. Then, they trained machine learning models to predict which claims should be prioritized for fact-checking. Recently, Abercrombie and Batista‑Navarro~\cite{abercrombie2020sentiment} published a literature review of 61 studies. These studies address the automatic analysis of sentiment and opinions expressed and the positions taken by speakers in parliamentary (and other legislative) debates on those topics.''

In addition to NLP for analyzing parliamentary data, other studies utilized network science for data mining parliamentary data.

In 2005, Porter et al.~\cite{porter2005network} investigated the US House of Representatives network of committees and subcommittees. They showed that network theory, combined with the analysis of roll-call votes using singular value decomposition, successfully uncovers political and organizational correlations between committees in the House without the need to incorporate other political information. 

Additionally, Dal Maso et al.~\cite{dal2014voting} conducted a network analysis study that analyzed the network of relations between parliament members according to their voting behavior. As a case study, Dal Maso et al. focused on the Chamber of Deputies of the Italian Parliament. They find sharp contrasts in the political debate, which does not imply a relevant structure based on established parties. In addition, they introduce a way to track the stability of the government coalition over time. This can discern the contribution of each member and the impact of its possible defection.

In 2020, an open-source tool was developed to inspect and explore the Israeli parliament\footnote{\url{https://github.com/SgtTepper/BetaKnessetWeb}}. The tool allows you to search for various topics by keywords, search for parliament members' votes and look at all parliament members' activities. This tool enables citizens to explore all the popular topics discussed in parliament sessions. It also allows them to check politicians' opinions and interests.

Lastly, a tool named LOCALVIEW~\cite{barari2023localview} was published in 2023. LOCALVIEW is a valuable tool for studying American local government policy-making due to its unprecedented scale and coverage. This data set consists of over 139,000 videos, as well as corresponding textual and audio transcripts of local government meetings publicly uploaded to YouTube. It covers 1,012 places and 2,861 distinct governments across the United States from 2006 to 2022. LOCALVIEW is a breakthrough in its ability to provide real-time access to regional policy-making, which has been difficult and expensive to study at scale. This data set can aid scholars, journalists, and other observers of regional politics and policies in exploring substantive phenomena of their interest. It covers a wide range of municipalities and counties. Moreover, its applicability extends beyond local policy-making. It has implications for the study of deliberative democracy, interpersonal communication, and intergroup dynamics along partisan, racial, geographical, or other dimensions. Overall, LOCALVIEW presents a powerful tool for studying American local government policy-making and communication between constituents and government officials.

\subsection{Time Series Analysis}
\label{sec:timeseries}

This study analyzed data from time series observations at a given time interval. This study uses offline Change Point Detection (CPD) algorithms. CPD is the problem of estimating the point at which statistical properties of a sequence of observations change. Over the years, several multiple-change-point search algorithms have been proposed to overcome this challenge~\cite{wambui2015power}. CPD is mainly used for two purposes: (a) to identify abnormal sequences along with a time series; and (b) to identify sudden changes in real-time.~\cite{truong2020selective}.

Detecting those change points is challenging for many applications in different areas, from finance~\cite{zeileis2010testing} to Bio-informatics~\cite{erdman2008fast}. For the first purpose, offline CPD algorithms are used. Offline CPD algorithms assume we have all the necessary data to process. As a result, the algorithms will find all the change points over time and not just the latest changes~\cite{truong2020selective}.

Our study utilized four popular offline CPD algorithms:

\begin{enumerate}

  \item \textit{Pruned Exact Linear Time (PELT)}- PELT detects change points through cost minimization~\cite{killick2012optimal}.

  \item \textit{The binary segmentation method (BinSeg)}- BinSeg is a sequential approach that works as follows: first, one change point is detected in the whole sequence, then, the series is split around the detected change point, and the operation is recursively repeated on the two resulting 

  sub-sequences~\cite{scott1974cluster}.

  \item\textit{ Window-sliding (WinSlid)}- WinSlid is a window-based search method that ``computes the discrepancy between two adjacent windows that move along with the signal $y$.'' When the two windows differ, the discrepancy will be high accordingly, implying a change point~\cite{truong2020selective}.

  \item \textit{Dynamic programming search method (DYNP)} - DYNP finds the (exact) minimum of the sum of costs by computing the cost of all sub-sequences of a given signal. It is called "dynamic programming" because the search over all possible segmentation is ordered using a dynamic programming approach~\cite{truong2020selective}.

\end{enumerate}

\section{Methods and Experiments}
\label{sec:methods_and_experiments}

Our study concludes the parliamentary function by analyzing and processing the meetings of the parliamentary committees. We analyze thousands of open protocol documents to understand how parliament works over time. In addition, another goal of the study is to develop a generic platform that allows for easy monitoring of various parliamentary bodies and how they change over time.

Namely, given a data set with $n$ text documents (protocols), in any language, containing information centralizing a parliamentary body's work from $k$ different parliamentary committees over $t$ years. We developed a code framework that analyzes each document and creates a database. We utilize this database to discover and understand how the parliamentary body has functioned over the years.

The following subsections provide an overview of developing a generic framework for collecting structured information from parliamentary bodies. Our method consists of the following steps (see Figure ~\ref{fig:framework_flow}): First, we collect parliamentary data from publicly available sources (see Section~\ref{sec:data_collection_and_preprocessing}). Afterward, we parse the collected data into structured data and extract features, and utilize these features to identify abnormal meetings (see Section~\ref{sec:data_parsing}). Next, we will transform the structured data into time series per feature to allow us to run CPD algorithms to analyze the data and identify changing trends in the parliamentary work (see Section~\ref{sec:data_analyzsis}).
Lastly, we will evaluate the various CPD algorithms' performances by consulting domain experts to label the detected CPDs manually (see Section~\ref{sec:evaluation}).

\subsection{Computational Framework for Analyzing Parliamentary Data}
\label{sec:framework}

\subsubsection{Data Collection and Preprocessing.}
\label{sec:data_collection_and_preprocessing}

We developed our framework to collect data from various parliaments using different methods, such as official APIs and web scraping applications, which extract the necessary data according to the settings for each source (see Section~\ref{sec:open-goverment-info}). After collecting the data, we preprocess the data and clean it. If the collected data contain corrupted or missing information, we will try to extract the missing data manually. For example, suppose some committee meetings are classified. In this case, the scrapper will fail to download the meeting files. Therefore, we will manually update basic features about the meetings, such as committee ID and meeting date. At the end of this step, we will have collected data stored in various formats, such as Word, PDF, CSV, and JSON files.

\subsubsection{Data Parsing and Feature Extraction.}
\label{sec:data_parsing}

Given the data we have collected, the next step is to take the unstructured data collected and turn it into structured data. For this purpose, we parsed the data by performing the following:

First, we developed a parser for each parliament that receives information about the parliament committees' meetings as input, collected in various formats. Our parsers primarily utilize public Python libraries to process different types of files. Additionally, we use tools like regular expressions to extract the values according to a shared schema we predefined (see Table \ref{table:general_committee_schema}). Second, to support the unique properties of the different parliaments, we constructed a ``custom'' schema that extends the general schema and enriches it with fields relevant only to the specific parliament (see Table \ref{table:us_committee_schema}).

Next, we extracted and selected generic features that characterize parliamentary meetings from the structured data. The feature selection process was done by examining which features are seen in most meetings. It also examined which features could indicate a particular activity trend by comparing the most common features and differences among parliaments. The features we extracted and selected are described in Table~\ref{table:extracted_features}.

\begin{table*}
\caption{General Committee Schema}
\label{table:general_committee_schema}
\centering
\begin{tabular}{|l|l|} 
\hline
\textbf{\textbf{General Features}} & \textbf{Explanation}                                                                    \\ 
\hline
\textit{Country}                   &  Country name                                                                        \\ 
\hline
\textit{Parliament name}           & Parliament's Name     \\ 
\hline
\textit{Parliament ID}         & Parliament ID                                    \\ 
\hline
\textit{Parliament type}           & Parliament's type                                         \\ 
\hline
\textit{Committee ID}              & Committee ID                                            \\ 
\hline
\textit{Committee name}            & Committee name                                           \\ 
\hline
\textit{Meeting ID}                & Meeting ID                                             \\ 
\hline
\textit{Title}                     & The meeting's title      \\ 
\hline
\textit{Date}                      & The date on which the meeting was held                                                \\ 
\hline
\textit{Committee members}         & A list of committee members who attended the meeting                                                     \\ 
\hline
\textit{Members of Parliament}     & A list of parliament members who attended the meeting                                        \\ 
\hline
\textit{Document length}           & The amount of characters in the meeting documentation  \\
\hline
\end{tabular}
\end{table*}

\begin{table*}[h]
\caption{US Committee Schema}
\label{table:us_committee_schema}
\centering
\begin{tabular}{|l|l|} 
\hline
\textbf{\textbf{General Features}} & \textbf{Explanation}                                   \\ 
\hline
\textit{Serial numbers}           & The serial numbers of the meeting                   \\ 
\hline
\textit{Witnesses}                   & A list of witness's names who attended the meeting                          \\ 
\hline
\textit{Location ID}               & The location identifier where the meeting occurred  \\
\hline
\end{tabular}
\end{table*}

\begin{table*}
\caption{Extracted Features}
\label{table:extracted_features}
\centering
\begin{tabular}{|l|l|} 
\hline
\textbf{\textbf{General Features}} & \textbf{Explanation}                                  \\ 
\hline
\textit{Committee ID}              & The committee ID                       \\ 
\hline
\textit{Meeting text length}            & The number of characters
in the meeting report        \\ 
\hline
\textit{Meeting Duration}            & The meeting duration
        \\ 
\hline
\textit{Year}                      & The year in which the meeting held                                          \\ 
\hline
\textit{Month}                     & The month in which the meeting held                                         \\ 
\hline
Number of committee members        & The number of committee members \\ & who attended the meeting   \\ 
\hline
Number of parliament members       & The number of parliament members \\
                                   & who attended the meeting  \\
\hline
\end{tabular}
\end{table*}

\subsubsection{Data Analysis.}
\label{sec:data_analyzsis}

After parsing the data, we transformed the parsed data in order to present information about similarities or differences between parliaments or various committees of the same parliament. Furthermore, we analyzed the transformed data to detect trends or anomalies. For this purpose, we examined the following features and trends over time:
\begin{itemize}
    \item \textit{Number of committee meetings}.
    \item \textit{Average number of committee members}.
    \item \textit{Average text length of a meeting}.
\end{itemize}

For each parliamentary committee (denoted $C$), such as the US Congress Energy and Commerce Committee, we calculate the distribution of each of the three features described above (denoted $f$).
We use these distributions to identify anomalous protocols based on the interquartile range (IQR). The interquartile range (IQR) is a measure of statistical dispersion, which is the spread of the data~\cite{schwertman2004simple}. It is defined as the difference between the 75$^{th}$ (denoted Q3) and 25$^{th}$ (denoted Q1) percentiles of the data. To detect outliers using this method, we define two values, a lower bound, and an upper bound, defined as follows:

\[IQR: Q3 - Q1 \]
\[Lower Bound: (Q1 - 1.5 \cdot IQR)\]
\[Upper Bound: (Q3 + 1.5 \cdot IQR). \]

Any value below the lower bound or higher than the upper bound is considered an outlier.
For example, we can calculate the distribution of the number of committee members in the Energy and Commerce Committee's meetings. Then, we can calculate the bounds and identify specific meetings with an odd number of attending committee members.

Moreover, for each committee $C$ and each feature $f$, we generated a time series $g^C_f(t)$, where $t$ is a period. In this study, we chose the values of $t$ to be one of three specific time resolutions: per month, three months, and six months.

Because each parliament has a different working schedule and manages its meetings differently, we needed to figure out the proper time resolution to represent our time series.

\begin{figure*}[!ht]
    \centering
    \scalebox{0.75}{\includegraphics{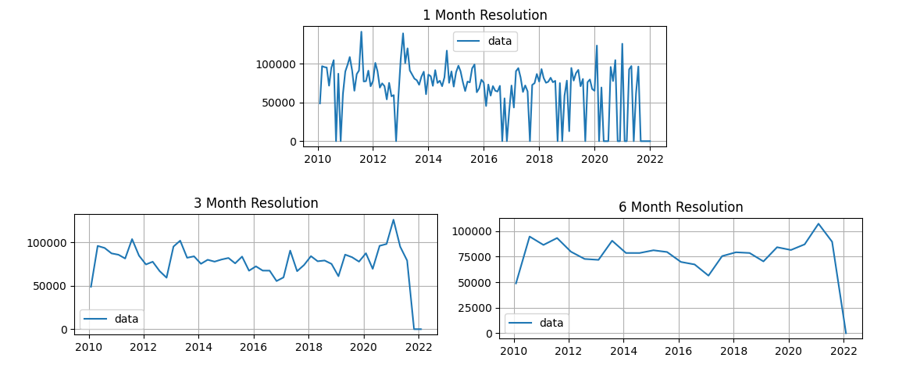}}
    \caption{Israel's Foreign Affairs Committee Protocols' Text Lengths over Time with Different Time Resolutions.}
    \label{fig:time_res_examine}
\end{figure*}

To overcome this challenge and find the adequate time resolution for representing our data, we plot graphs for each feature and committee in various time resolutions and charts that compare multiple time series (see Figure~\ref{fig:time_res_examine}).

Afterward, by observing the variety of generated time series with different time resolutions for each country, we manually selected the resolution that created a suitable time series. We then created three time series for each committee using the appropriate time resolution, one for each feature. On each time series, we run the following CPD algorithms: PELT, BinSeg, Window-based, and DYNP (see Section~\ref{sec:timeseries}). By manually inspecting each algorithm's results for several committees, we will choose two algorithms that perform most adequately.

\subsubsection{Evaluation.}
\label{sec:evaluation}

To evaluate our framework's performance on the different committees, we performed the following evaluations: First, we manually checked the meetings with abnormal values according to the IQR thresholds and calculated true and false positive rates according to the meetings' properties. We evaluated whether the IQR identified anomalous meetings.

Second, to evaluate CPD algorithms' performance. We ran the two selected CPD algorithms on the generated time series. Then, for each detected case, we assessed by consulting experts if it was a true or false event. Lastly, we could not evaluate the model using the accuracy or recall metrics because we are unaware of the points (events) we missed detecting, so we assess each algorithm's performance using \textit{Precision} and \textit{False Positive Rate (FPR)} defined as follow~\cite{ferri2009experimental}:

\[Precision: \frac{TP}{TP + FP} \]
\[FPR: \frac{FP}{TP + FP} \]

\subsection{Experimental Setup}
\label{sec:experimentals_setup}

\subsubsection{Data collection}
\label{sec:data_collection}

To evaluate our framework and algorithms, we collected parliamentary data from three countries:

\begin{itemize}
    \item \textbf{United States} - In the US, we collected the committees' data of the United States House of Representatives by scraping the US government's official website.\savefootnote{us_govinfo_footnote} {\url{https://www.govinfo.gov/}} We collected all available committees' data from January 1999 to the May of 2021 and parsed them into the framework's schema described in Table~\ref{table:general_committee_schema}. In addition, the scrapper can be configured to collect Senate committee data.
    
    \item \textbf{Canada} - In Canada (CA), we collected data from Canada's House of Commons, the lower house of the Parliament of Canada. We gathered the committees' data by scraping the House of Commons official website.\savefootnote{ca_scrapper_footnote}{\url{https://www.ourcommons.ca/en}} We collected all the available committees' data from January 2009 to May 2022 and parsed them into the framework's schema described in Table~\ref{table:general_committee_schema}.
    
    \item \textbf{Israel} - The Israeli parliament, The Knesset, offers all the committees' reports on their official website.\savefootnote{knesset_footnote}{\url{https://main.knesset.gov.il/Activity/committees/Pages/AllCommitteeProtocols.aspx}}
    In this study, we obtained a dataset with these reports from the different Knesset committees from January 1999 to January 2021.\footnote{To make this study reproducible, we also developed a web crawler that can download all these reports.} Next, we analyze those reports and parse them into the framework's schema described in Table~\ref{table:general_committee_schema}.
    
\end{itemize}

After parsing all the data collected for each country, we created time series only for the top 10 most active committees with the highest number of meetings, and that was our curated dataset.

\subsubsection{Features Distributions.}

For each country, we looked closely at each country's top-3 most active committees. Namely, we calculated the features' distributions for each country's top-3 most active committees and the three features. Using the IQR outlier method, we identified meetings with anomalous features (see Section~\ref{sec:data_analyzsis}). Then, we manually examined randomly selected abnormal meetings to determine whether they were genuinely anomalous. Finally, we evaluate this approach's true-positive rates for detecting abnormal features' values.

\subsubsection{Change Point Detection.}
\label{sec:cpd_experiments}

We focused on the Israeli parliament to evaluate CPD algorithms' performance. As described in Section~\ref{sec:data_collection_and_preprocessing}, we created time series only for the ten most popular committees, and for each committee, we generated three different time series, one for each feature. Next, we examined our data and found that the suitable time resolution to represent our data was six months of aggregation.

We manually inspected the four CPD algorithm results and found that the PELT and the DYNP algorithms presented adequate results in detecting CPDs. Therefore, we focused on evaluating the PELT and DYNP algorithms.

To accurately evaluate the performance of the PELT and the DYNP algorithms, we needed to check that the change points detected by the algorithms were actual events and not false positive detection.
In order to detect actual events around those dates, we conduct a comprehensive investigation.

We checked and cross-referenced various Knesset systems, applications, databases, protocols, and records. Moreover, if possible, we interviewed committee administrators, committee members, and veteran committee directors and asked them what happened at those times. They checked and gave us answers and explanations for cases they knew about.

The classifications range from ``no justifiable reason'' to ``Knesset summer recess/vacation days'' and ``COVID-19 days and special discussions in specific committees'' (see Figure~\ref{fig:cpd_example}).

\section{Results}
\label{sec:results}

As described in Section~\ref{sec:data_collection}, using dedicated web crawlers, we collected committees’ data from three countries (see Table~\ref{table:collected_data}). Overall, we collected the following data: from the US parliament, 16,989 protocols from 21 committees for 22 years. We collected 12,124 protocols from the Canadian parliament from 44 committees for 13 years. Lastly, we downloaded 35,800 protocols from 33 committees for the Israeli parliament for 22 years.

\begin{table*}[ht]
\centering
\caption{Collected Parliamentary Data Overview}
\label{table:collected_data}
\begin{tabular}{|l|c|c|c|} 
\toprule
\textbf{Country} & \textbf{Number of committees} & \textbf{Number of protocols} & \textbf{Time period}  \\ 
\hline
US               & 21                            & 16,989                       & 01/1999--05/2021             \\ 
\hline
CA           & 44                            & 12,124                       & 01/2009--05/2022             \\ 
\hline
IL           & 33                            & 35,800                       & 01/1999–01/2021             \\
\bottomrule
\end{tabular}
\end{table*}

By analyzing the number of collected protocols for each country, we observed that the average and median number of protocols per committee in the US is 809. The average number of protocols per committee in CA is 275, and the median is 416. Lastly, Israel's average number of protocols per committee is 1,084.8, and the median is 226. 

\begin{figure*}[htp]

\centering\begin{subfigure}[b]{0.5\linewidth} 
\centering\includegraphics[width=6cm]{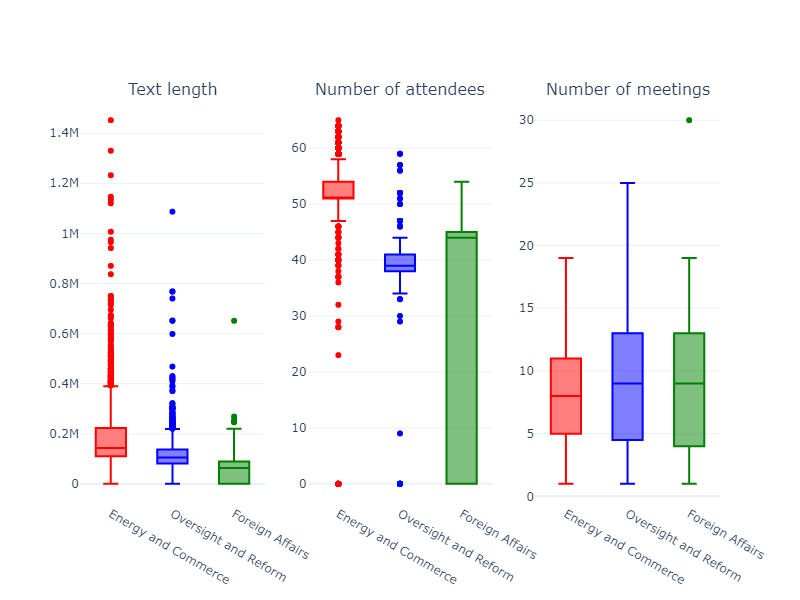} 
\caption{\label{fig:US 8010}US committees feature distributions.} 
\end{subfigure}\hfill
\begin{subfigure}[b]{0.5\linewidth} 
\centering\includegraphics[width=6cm]{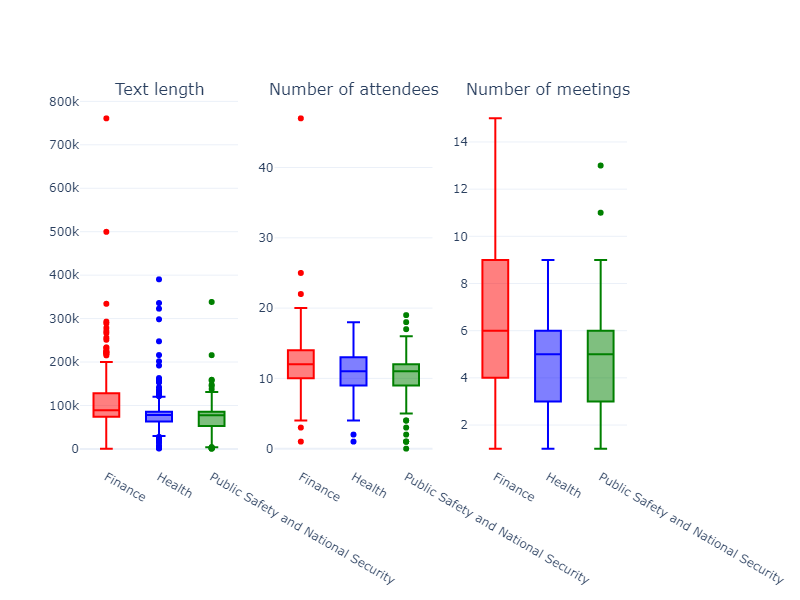} 
\caption{\label{fig:CA}CA committees feature distributions.} 
\end{subfigure}\vspace{10pt}

\begin{subfigure}[b]{\linewidth} 
\centering\includegraphics[width=6cm]{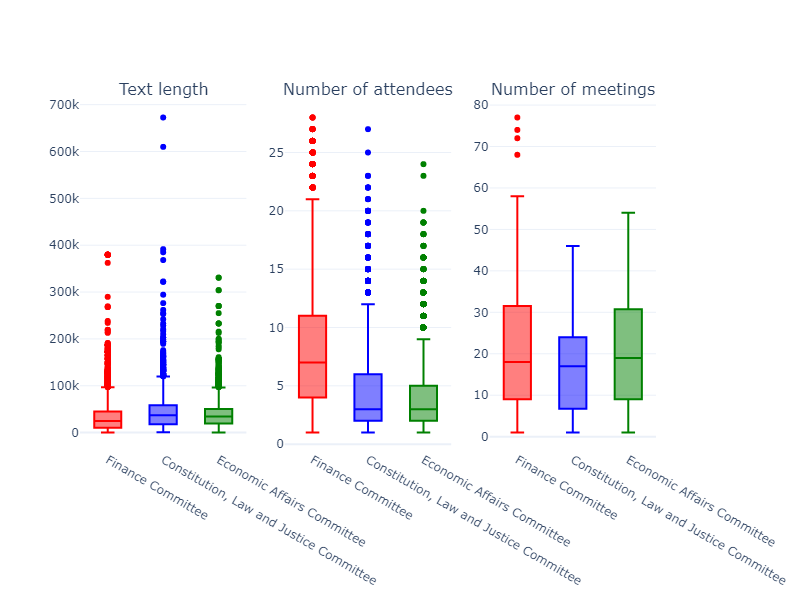} 
\caption{\label{fig:IL} IL committees feature distributions.} 
\end{subfigure} 
\caption{Countries' most active committees feature distributions.} 
\label{fig:countries_features}
\end{figure*}

Next, as described in Section~\ref{sec:data_parsing}, we analyzed the collected protocols and constructed a detailed database with each committee's meeting details (see Tables~\ref{table:general_committee_schema} and~\ref{table:extracted_features}). By utilizing the database, for each committee in each country, we calculated the committee's number of meetings, the average number of committee members, and the meetings' protocols' average text lengths (see Table~\ref{table:committees_types_meetings_per_country} and Figure~\ref{fig:countries_features}).

Moreover, we focused on the top-3 most active committees. We used the IQR method to uncover anomalous meetings. The IQR algorithm identified 656 anomalous protocols according to three features that were examined. We randomly sampled 263 of these protocols. For the average text length of a meeting feature, we randomly sampled 50 abnormal protocols from each county. For the average number of committee members feature, we randomly sampled 50 abnormal protocols from the USA and Israel. We also randomly selected six anomalous protocols from Canada.

For the number of committee meetings feature, which contained only a few anomalous protocols, we selected one, two, and four protocols for the USA, Canada, and Israel, respectively. Then, we manually examined these sampled protocols and observed that all recognized protocols were indeed abnormal and not a result of problems with the protocols' formats or issues with the parsing of these files.\footnote{In some cases, we noticed small gaps between the features' values calculated by our algorithms and the features observed in the manual analysis of the protocols. Nevertheless, in most cases, the gaps were relatively minor. For example, we observed slight differences in the protocols' number of characters or minor differences in the number of members' attendees.} Using this method, we uncover interesting and abnormal meetings: For example, we identified a charged meeting of the Finance Committee in the Canadian Parliament\footnote{\url{https://www.ourcommons.ca/DocumentViewer/en/41-1/FINA/meeting-94/evidence}} which discussed the implementation of sections of the budget and was broadcast on TV. In this meeting, the committee chair and members requested the Finance Minister's presence and not officials on his behalf. Many people were in this discussion, and the meeting was relatively long. An additional example, the IQR algorithm indicated a hearing titled "BP's Role in the Deepwater Horizon Explosion and Oil Spill" held by the US House Energy and Commerce Committee, the Oversight and Investigations subcommittee. This hearing included a long discussion with many people present.

\begin{table*}
\centering
\caption{The Committees with the Highest Number of Meetings per Country}
\label{table:committees_types_meetings_per_country}
\begin{tabular}{|l|c|c|} 
\hline
{\textbf{Country}} & {\textbf{Committee Name}}                & \textbf{{Number of Meetings}}  \\ 
\hline
IL                       & Immigration, Absorption and Diaspora  Affairs  & 1563                             \\ 
\cline{2-3}
                         & The Status of Women and Gender Equality        & 1302                             \\ 
\cline{2-3}
                         & Constitution, Law and Justice                  & 3973                             \\ 
\cline{2-3}
                         & Economic Affairs                               & 4618                             \\ 
\cline{2-3}
                         & Education, Culture and Sports                  & 3460                             \\ 
\cline{2-3}
                         & Finance                                        & 5530                             \\ 
\cline{2-3}
                         & House                                          & 2065                             \\ 
\cline{2-3}
                         & Internal Affairs and Environment               & 3501                             \\ 
\cline{2-3}
                         & Labor and Welfare                              & 3918                             \\ 
\cline{2-3}
                         & State Control                                  & 1665                             \\ 
\hline
CA                       & Agriculture and Agri-Food                      & 481                              \\ 
\cline{2-3}
                         & Finance                                        & 736                              \\ 
\cline{2-3}
                         & Government Operations and Estimates            & 511                              \\ 
\cline{2-3}
                         & Health                                         & 520                              \\ 
\cline{2-3}
                         & Human Resources, Skills and Social Development & 507                              \\ 
\cline{2-3}
                         & Industry, Science and Technology               & 487                              \\ 
\cline{2-3}
                         & Justice and Human Rights                       & 487                              \\ 
\cline{2-3}
                         & Procedure and House Affairs                    & 508                              \\ 
\cline{2-3}
                         & Public Accounts                                & 474                              \\ 
\cline{2-3}
                         & Public Safety and National Security            & 516                              \\ 
\hline
US                       & Small Business                                 & 904                              \\ 
\cline{2-3}
                         & Armed Services                                 & 1040                             \\ 
\cline{2-3}
                         & Science, Space, and Technology                 & 848                              \\ 
\cline{2-3}
                         & Energy and Commerce                            & 1692                             \\ 
\cline{2-3}
                         & Financial Services                             & 1265                             \\ 
\cline{2-3}
                         & Foreign Affairs                                & 1516                             \\ 
\cline{2-3}
                         & Homeland Security and Governmental Affairs     & 821                              \\ 
\cline{2-3}
                         & Oversight and Reform                           & 2220                             \\ 
\cline{2-3}
                         & Natural Resources                              & 929                              \\ 
\cline{2-3}
                         & the Judiciary                                  & 1130                             \\
\hline
\end{tabular}
\end{table*}

Lastly, we used the PELT and DYNP CPD algorithms and detected events in the Israeli Knesset. We applied the two algorithms with six months of time resolution to all three features.

\begin{figure*}
    \centering
    \scalebox{0.75}{\includegraphics{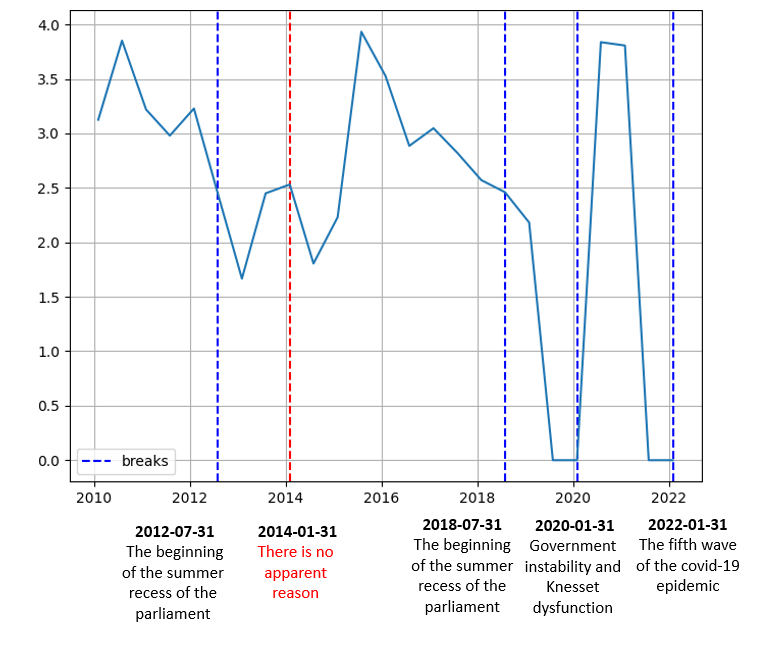}}
    \caption{{DYNP detected CPDs on the Knesset's Committee for Immigration, Absorption, and Diaspora Affairs  - blue horizontal lines mark true positive events, and the red horizontal line marks a false positive detected event.}}
    \label{fig:cpd_example}
\end{figure*}

The two algorithms returned about 289 change points; 124 and 165 change points were detected by the PELT and DYNP algorithms, respectively. We then performed a comprehensive investigation (see Section~\ref{sec:cpd_experiments}) and manually classified each detected point as a true or a false event, and if possible, we added the event's reason. For example, we identified events due to Knesset summer recess, COVID-19 days, and special discussions in specific committees or government instability and dysfunction of the Knesset (see Figure~\ref{fig:cpd_example}). Both algorithms often produced accurate results with perfect precision of 1.

Overall, precision-wise, the DYNP algorithm achieved better results than the PELT algorithm, with an average precision of 0.76 vs. 0.604 (see Table~\ref{table:cpd_precision_fpr_scores}).

\section{Discussion}
\label{sec:diss}

By analyzing the results presented in Section~\ref{sec:results}, the following can be noted:

First, a vast amount of government data is accessible online. Using different data collection methods described in Section~\ref{sec:open-goverment-info}, we utilized the CAPD framework to collect more than 64,000 government meeting protocols from three countries over more than twenty years (see Table~\ref{table:collected_data}). This accumulated unstructured data corpus can help us study various government activities in different geographical locations over extended periods. 

Second, even though the collected data were from different countries, using our framework, it was feasible to convert the collected unstructured data into structured data (see Section~\ref{sec:data_parsing} and Tables~\ref{table:general_committee_schema},~\ref{table:us_committee_schema}, and ~\ref{table:extracted_features}). Converting the data into structured data makes it possible to use the proposed framework to gather data from various countries. This makes it possible to compare countries' activities to specific events, such as climate change or emergency. 

Third, we demonstrate that it is possible to use data science tools to analyze the collected protocols to monitor governments' committees' activities over time. Making this type of analysis open to the public can help make governments more transparent. Moreover, it can also help government members monitor their activities. 

Fourth, in this study, we demonstrated that it is possible to analyze the structured parliamentary data with anomaly detection algorithms, such as IQR, to identify abnormal meetings, such as anomalous meetings with many participants or extensive duration (see Section~\ref{sec:results}). Moreover, we can utilize this algorithm to detect periods in time with a relatively high number of committee meetings. Using this method, it is possible to point out interesting meetings or meetings with public interest, such as the ``BP's Role in the Deepwater Horizon Explosion and Oil Spill'' hearing (see Section~\ref{sec:results}).

Fifth, we showed that it is possible to utilize CPD algorithms, such as the PELT and the DYNP algorithms (see Section~\ref{sec:timeseries}), to detect actual events that affected the parliamentary activities of different committees, such as summer breaks or the COVID-19 pandemic (see Figure~\ref{fig:cpd_example}). We demonstrated that with some parameter algorithm tuning, such as time resolution tuning, it is possible to use CPD algorithms to identify events with relatively high precision. However, as shown in Table~\ref{table:cpd_precision_fpr_scores}, different CPD algorithms can work differently and return different results. For example, the DYNP algorithm produced better results than the PELT algorithm. There are several cases where the PELT algorithm demonstrated higher precision than DYNP. Therefore, there is a need to develop a method to select the best-performing CPD algorithm for each committee. It is our hope to investigate this in future research.

Lastly, the methods and framework presented in this study have several limitations worth noting. First, as we demonstrated, the framework works in different parliaments. It still needs some manual adjustments to collect data from other parliaments. Second, different parliaments work differently and at different times of the year. Therefore, utilizing the various algorithms, such as IQR and CPD algorithms, may also require manual adjustment. This includes time resolution setting and adjusting features calculations. Nevertheless, we developed our framework to be generic enough to support its extension to other parliaments.

\begin{table*}[hbt]
\centering
\caption{CPD Precision and FPR scores - IL}
\label{table:cpd_precision_fpr_scores}
\arrayrulecolor{black}
\begin{tabular}{|l|l|l|l|l|l|} 
\hline
\textbf{Committee Name}                                                                 & \textbf{{Data }Type}   & \multicolumn{2}{l|}{\textbf{Precision}}          & \multicolumn{2}{l|}{\textbf{FPR}}                                                     \\
                                                                                        &                              & \textbf{\textbf{PELT~}} & \textbf{\textbf{DYNP}} & \textbf{\textbf{\textbf{\textbf{PELT~}}}} & \textbf{\textbf{\textbf{\textbf{DYNP}}}}  \\ 
\hline
\begin{tabular}[c]{@{}l@{}}Immigration, Absorption, \\and Diaspora Affairs\end{tabular} & Number of meetings           & 0.6                     & \textbf{1}             & 0.4                                       & \textbf{0}                                \\ 
\cline{2-6}
                                                                                        & Number of members            & 0.5                     & \textbf{0.8}           & 0.5                                       & \textbf{0.2}                              \\ 
\cline{2-6}
                                                                                        & Text Length                  & 0.6                     & 0.6                    & 0.4                                       & 0.4                                       \\ 
\hline
\begin{tabular}[c]{@{}l@{}}Status of women,\\~and gender equality\end{tabular}          & Number of meetings           & 0.75                    & \textbf{0.8}           & 0.25                                      & \textbf{0.2}                              \\ 
\cline{2-6}
                                                                                        & Number of members            & 0.33                    & \textbf{0.8}           & 0.67                                      & \textbf{0.2}                              \\ 
\cline{2-6}
                                                                                        & Text Length                  & 0.6                     & \textbf{0.8}           & 0.4                                       & \textbf{0.2}                              \\ 
\hline
\begin{tabular}[c]{@{}l@{}}Constitution, Law,\\~and Justice~\end{tabular}               & Number of meetings           & \textbf{0.75}           & 0.6                    & \textbf{0.25}                             & 0.4                                       \\ 
\cline{2-6}
                                                                                        & Number of members            & 0.67                    & \textbf{0.8}           & 0.33                                      & \textbf{0.2}                              \\ 
\cline{2-6}
                                                                                        & Text Length                  & 0.6                     & 0.6                    & 0.4                                       & 0.4                                       \\ 
\hline
Economic Affairs                                                                        & Number of meetings           & 0.6                     & \textbf{0.8}           & 0.4                                       & \textbf{0.2}                              \\ 
\cline{2-6}
                                                                                        & Number of members            & 0.5                     & \textbf{0.8}           & 0.5                                       & \textbf{0.2}                              \\ 
\cline{2-6}
                                                                                        & Text Length                  & 0.6                     & 0.6                    & 0.4                                       & 0.4                                       \\ 
\hline
\begin{tabular}[c]{@{}l@{}}Education, Culture,\\~and Sports\end{tabular}                & Number of meetings           & 0.67                    & \textbf{0.8}           & 0.33                                      & \textbf{0.2}                              \\ 
\cline{2-6}
                                                                                        & Number of members            & 0.5                     & \textbf{0.8}           & 0.5                                       & \textbf{0.2}                              \\ 
\cline{2-6}
                                                                                        & Text Length                  & 0.6                     & \textbf{0.8}           & 0.4                                       & \textbf{0.2}                              \\ 
\hline
Finance                                                                                 & Number of meetings           & 0.5                     & \textbf{0.8}           & 0.5                                       & \textbf{0.2}                              \\ 
\cline{2-6}
                                                                                        & Number of members            & 0.5                     & \textbf{0.8}           & 0.5                                       & \textbf{0.2}                              \\ 
\cline{2-6}
                                                                                        & Text Length                  & 0.6                     & \textbf{0.8}           & 0.4                                       & \textbf{0.2}                              \\ 
\hline
House~                                                                                  & Number of meetings           & 0.5                     & \textbf{0.8}           & 0.5                                       & \textbf{0.2}                              \\ 
\cline{2-6}
                                                                                        & Number of members            & 0.5                     & \textbf{0.8}           & 0.5                                       & \textbf{0.2}                              \\ 
\cline{2-6}
                                                                                        & Text Length                  & 0.6                     & 0.6                    & 0.4                                       & 0.4                                       \\ 
\hline
\begin{tabular}[c]{@{}l@{}}Internal Affairs,\\~and Environment~\end{tabular}            & Number of meetings           & \textbf{1}              & 0.6                    & \textbf{0}                                & 0.4                                       \\ 
\cline{2-6}
                                                                                        & Number of members            & \textbf{1}              & 0.6                    & \textbf{0}                                & 0.4                                       \\ 
\cline{2-6}
                                                                                        & Text Length                  & 0.6                     & 0.6                    & 0.4                                       & 0.4                                       \\ 
\hline
Laber and Welfare~                                                                      & Number of meetings           & 0.75                    & \textbf{0.8}           & 0.25                                      & \textbf{0.2}                              \\ 
\cline{2-6}
                                                                                        & Number of members            & 0.5                     & \textbf{0.6}           & 0.5                                       & \textbf{0.4}                              \\ 
\cline{2-6}
                                                                                        & Text Length                  & 0.6                     & \textbf{0.8}           & 0.4                                       & \textbf{0.2}                              \\ 
\hline
State Control~                                                                          & Number of meetings           & 0.67                    & \textbf{1}             & 0.33                                      & \textbf{0}                                \\ 
\cline{2-6}
                                                                                        & Number of members            & 0.33                    & \textbf{1}             & 0.67                                      & \textbf{0}                                \\ 
\cline{2-6}
                                                                                        & Text Length                  & 0.6                     & \textbf{0.8}           & 0.4                                       & \textbf{0.2}                              \\ 
\hline
                                                                                  \cellcolor{lightgray}      & \cellcolor{lightgray}\textbf{Number of meetings }        & \cellcolor{lightgray}0.679                   & \cellcolor{lightgray}\textbf{0.8}           & \cellcolor{lightgray}0.321                            & \cellcolor{lightgray}\textbf{0.2}                                       \\ 
\cline{2-6}

\textbf{ \cellcolor{lightgray}{Average Score}}                                                          & \cellcolor{lightgray}\textbf{Number of members}          &  \cellcolor{lightgray}0.533                   & \cellcolor{lightgray}\textbf{0.78}          & \cellcolor{lightgray}0.467                            & \cellcolor{lightgray}\textbf{0.22}                                      \\ 

\cline{2-6}
                                                              \cellcolor{lightgray}                          & \cellcolor{lightgray}\textbf{Text Length}                & \cellcolor{lightgray}0.6                     & \cellcolor{lightgray}\textbf{0.7}           & \cellcolor{lightgray}0.4                              & \cellcolor{lightgray}\textbf{0.3}                                       \\
\hline
\end{tabular}
\end{table*}

\section{Conclusions}
\label{sec:conclusions}

This study presents the CAPD framework, an open and generic framework that enables collecting and analyzing large-scale parliamentary data from multiple sources. The framework can collect texts of parliamentary protocols, such as committee meeting protocols, and parse them into structured data (see Section~\ref{sec:methods_and_experiments} and Figure~\ref{fig:framework_flow}), and analyze the data using state-of-the-art data science algorithms and tools. To test and evaluate the CAPD framework, we collected 64,913 protocols, from 264 committees, from three countries (see Section~\ref{sec:results}). We parsed these protocols using the framework. We calculated different committees' features, and generated a time series illustrating the various committees' activities over long periods. Then, we utilized IQR to uncover abnormal events and CPD algorithms to detect events that influenced committees' activities (see Section~\ref{sec:methods_and_experiments}), such as the COVID-19 pandemic. Lastly, we evaluate the detected CPDs, which indicates that CPD algorithms performed adequately with relatively high precision rates.

This study has several future research directions: First, we can extend the framework's collection capabilities and collect more data from more countries and other parliamentary bodies. Second, we can calculate additional features from the protocols, such as the number of speakers and speakers' average word number. Moreover, we can use Natural Language Processing (NLP) algorithms and tools, such as BERT Topics~\cite{grootendorst2022bertopic} and various sentiment analysis algorithms~\cite{birjali2021comprehensive} to analyze the protocols and extract additional features regarding the topics and content of each protocol. Third, we can develop an interactive website that, based on CAPD framework analysis, presents different real-time features on other committees. The website can help the public monitor government activities better.

Lastly, the framework can be extended to collect data from different countries on specific topics. This data collection will enable comparative studies regarding governments' performance in handling global events, such as climate change and COVID-19.

Overall, the CAPD framework can harness worldwide open parliamentary data to advance government research and monitoring, promoting research in the field of public policy and helping make governments more transparent.

\section{Code and Data Availability}
The CAPD framework's code and all the collected data will be available upon publication

\bibliographystyle{unsrt}
\bibliography{main.bib}


\end{document}